\documentclass[reprint,aps,pra,showpacs,amsmath,amssymb,twocolumn,floatfix,10pt,groupedaddress,longbibliography]{revtex4-1}
\pdfoutput=1
\usepackage[export]{adjustbox}
\usepackage{graphicx}
\usepackage{dcolumn}
\usepackage{bbold}
\usepackage{bm}
\usepackage{epstopdf}
\usepackage{braket}
\usepackage{mathtools}
\usepackage[svgnames]{xcolor}
\PassOptionsToPackage{hyphens}{url}
\usepackage{hyperref}
\hypersetup{colorlinks=true, citecolor=NavyBlue, urlcolor=NavyBlue, linkcolor=NavyBlue}
\usepackage[caption=false]{subfig}

\newcommand{\figref}[1]{Fig.~\ref{#1}}
\usepackage{etoolbox}
\apptocmd{\sloppy}{\hbadness 10000\relax}{}{}

\begin{document}

\title{Nonlocal Generalized Quantum Measurements\texorpdfstring{\\ of Spin Products Without Maximal Entanglement}{}}

\author{Pierre Vidil}
\email{pierre@quantum.riec.tohoku.ac.jp}
\author{Keiichi Edamatsu}
\affiliation{Research Institute of Electrical Communication, Tohoku University, Sendai 980-8577, Japan}

\date{\today}

\begin{abstract}

Measuring a nonlocal observable on a space-like separated quantum system is a resource-hungry and experimentally challenging task. 
Several theoretical measurement schemes have already been proposed to increase its feasibility, using a shared maximally-entangled ancilla. 
We present a new approach to this problem, using the language of generalized quantum measurements, to show that it is actually possible to measure a nonlocal spin product observable without necessarily requiring a maximally-entangled ancilla. 
This approach opens the door to more economical arbitrary-strength nonlocal measurements, with applications ranging from nonlocal weak values to possible new tests of Bell inequalities. 
The relation between measurement strength and the amount of ancillary entanglement needed is made explicit, bringing a new perspective on the links that tie quantum nonlocality, entanglement and information transmission together.

\end{abstract}

\maketitle

\section{Introduction}

\addtocounter{figure}{1}
\begin{figure*}
    \subfloat{
      \label{sfig:erasure}
      \includegraphics[scale=0.23,valign=c]{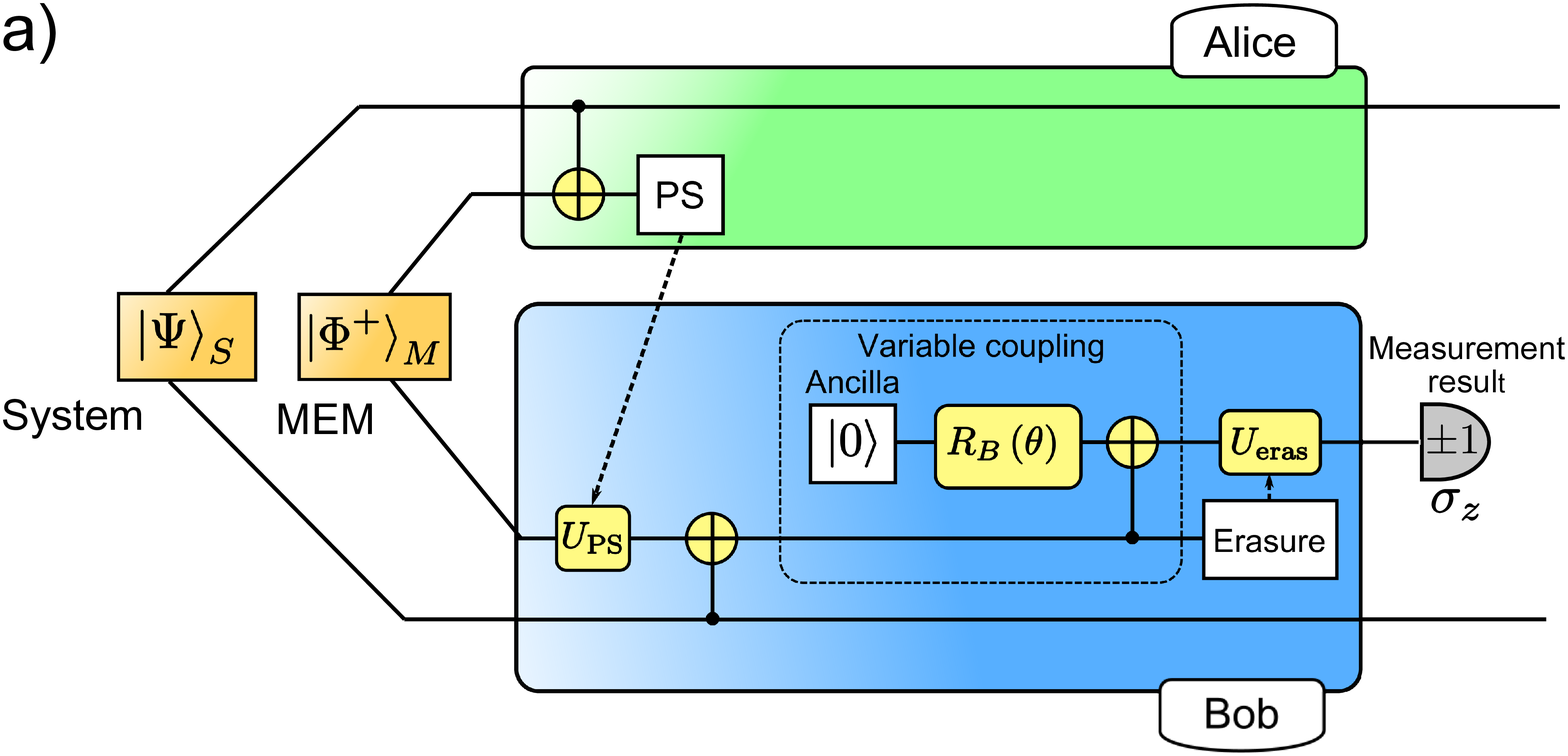}%
    }\hfill
    \subfloat{
      \label{sfig:nmem}
      \includegraphics[scale=0.23,valign=c]{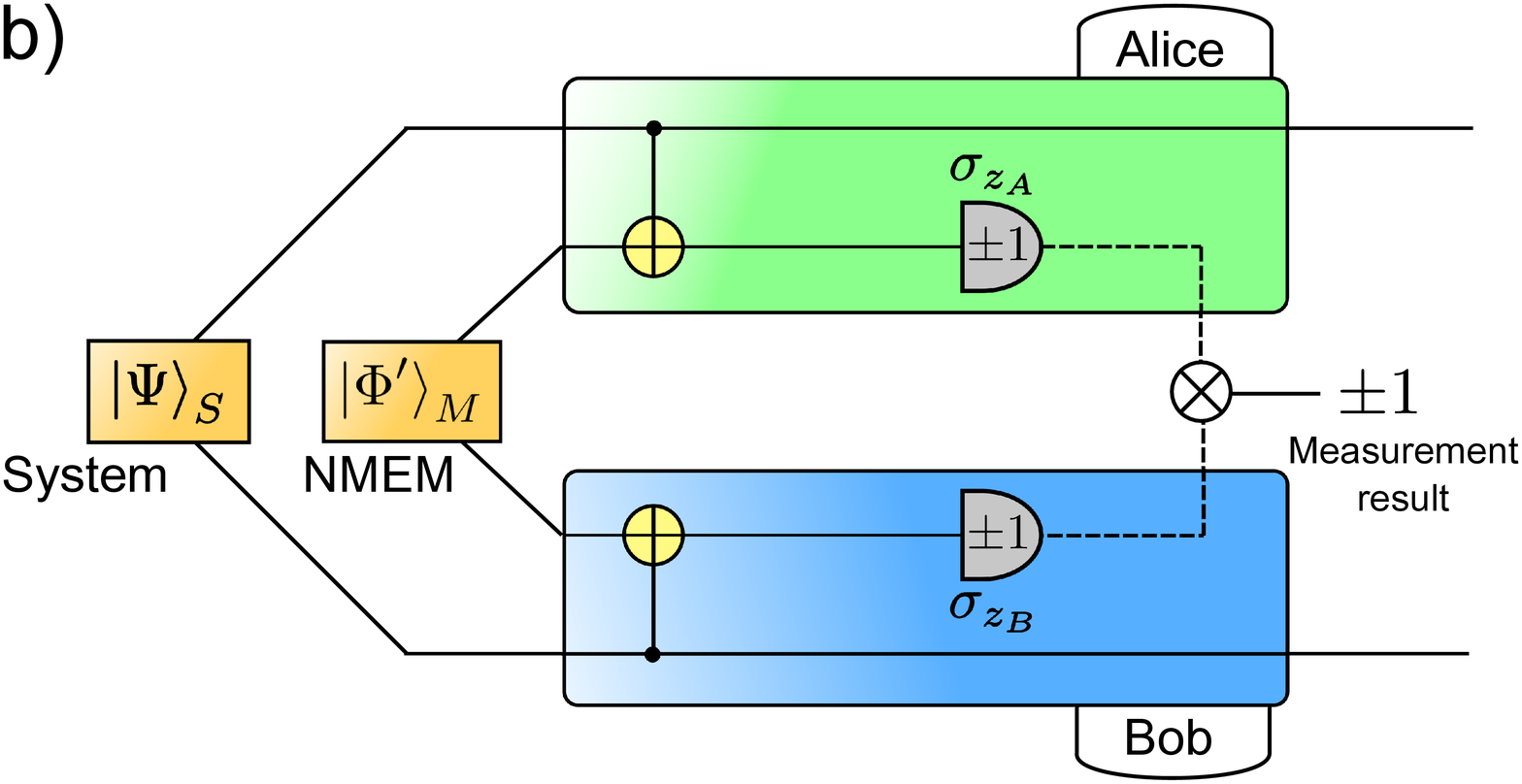}%
    }
    \caption{Comparison between the quantum circuit representations of the quantum erasure scheme (a) proposed in Ref. \cite{Brodutch2016} and this Paper's approach (b) in the case of a spin product measurement.
    The quantum erasure method uses a maximally-entangled meter (MEM), post-selection (PS) and an additional erasure step.
    In our method, Alice and Bob each apply a CNOT interaction between their system qubits and their shared non-maximmaly entangled meter (NMEM). 
    The global measurement result is computed after the local outcomes are reunited and multiplied.
    This presents a clear advantage in terms of practicality in that it does not necessarily require maximal entanglement and can be achieved with only four qubits.}
    \label{fig:compar}
\end{figure*}
\addtocounter{figure}{-2}

Almost since its inception, the behavior of space-like separated quantum systems has been at the heart of multiple heated controversies around quatum mechanics \cite{Einstein1935,Bell1964,Laloe2012}, as well as the key to some of its most promising technological applications. 
These include superdense coding \cite{Bennett1992,Mattle1996}, quantum teleportation \cite{Bennett1993,Bouwmeester1997}, entanglement swapping \cite{Zukowski1993,Pan1998} and device-independent quantum key distribution \cite{Barrett2005,Acin2006,Acin2007} among others. 
All have in common that they rely on the measurement of an operator that contains information about not just one, but several, possibly entangled, quantum particles.

Sometimes, one might be faced with a situation where those different parts are space-like separated and direct interaction between them is not available. 
The question of whether or not it is possible to measure such multipartite observables instantaneously in this case was first answered in the negative by Landau and Peierls \cite{Landau1931} in 1931, on the grounds of locality constraints. 

Yet it was proven much later that such \emph{nonlocal measurements} are in fact possible for certain observables, given adequate resources \cite{Aharonov1981,Aharonov1986,Popescu1994}. 
When the different parts are separated, they are made to strongly interact with an additional maximally-entangled state, a precious resource in quantum information \cite{Bennett1998,Horodecki2007}, that is used to carry out the measurement and store the result. 

This type of measurement scheme is often referred to as a \emph{von Neumann (VN) measurement} \cite{vonNeumann1935}, and the use of a maximally-entangled meter state has been shown to solve the problem of achieving complete Bell State Measurement \cite{Bennett1996,Beckman2001}, even in linear-optical systems \cite{Edamatsu2016a}.
The interaction between the system and the meter leading to the final result can then be made instantaneous, even though retrieving said result from the entangled meter requires some finite amount of time, as dictated by special relativity. 
Such a  strong VN measurement of nonlocal variables has already been implemented using hyperentangled photonic quantum systems \cite{Xu2019}.
Furthermore, if one is ready to part with the VN approach and discard the final state of the system, all nonlocal observables become measurable \cite{Groisman2002, Vaidman2003, Clark2010} via so-called verification measurements and finite entanglement consumption.

However VN measurements can be more than just strong (projective) measurements \cite{Wiseman2009}, which have been discussed so far. 
By suitably tailoring the system-meter interaction, as in \figref{fig:one-qubit}, one can manage to only retrieve part of the information about a quantum state, in order to somewhat preserve it \cite{Aharonov1987}.
This has been successfully applied to local systems for quantum metrology \cite{Hosten2008,Dixon2009}, or in quantum foundations when one wishes to limit the effects of the measurement back-action via \emph{weak measurements} \cite{Aharonov1988,Aharonov2002}.

One can naturally wonder if this type of interaction tuning can be extended to the nonlocal case.
The quantum erasure scheme, developed by Brodutch and Cohen \cite{Brodutch2016} and recently implemented by Li and al. \cite{Li2019}, provides a solution by effectively reproducing a nonlocal aribitrary-strength VN interaction. 
It also extends the class of measurable nonlocal observables, by inserting a probabilistic element that prevents running afoul of causality.

This comes at a price however: on top of a maximally-entangled meter, an extra local meter is necessary to store the result, thus making this method difficult to implement experimentally. 
The simplest case indeed requires a total of five distinct qubits as can be seen on \figref{sfig:erasure}.

In this Paper, we present a simpler method that can be used to measure nonlocal spin products, yielding the same post-measurement state evolution and statistics as the quantum erasure method, while using less resources, as shown in \figref{sfig:nmem}, where the total number of qubits necessary is four.
Our approach presents a complemetary point of view to the problem of nonlocal measurements that relies on the language of generalized quantum measurements \cite{Wiseman2009,Nielsen2010} applied to spin product observables.
We prove that in this particular case, it is possible to reproduce the behavior of an arbitrary-strength nonlocal measurement using a non-maximally-entangled meter, a weaker resource than what was needed in previous schemes.
In particular, we show that the optimal amount of meter entanglement necessary is directly related to the desired measurement strength, and that excessive entanglement may on the contrary degrade the purity of the post-measurement system state. 
One can then achieve a nonlocal weak measurement with only a limited amount of ancillary entanglement, which greatly increases experimental feasibility, notably for linear-optical implementations.

The structure of this Paper is as follows.
In Sec. \ref{sec:one-qubit}, we review one-qubit generalized measurements, which constitute the starting point for the later extension to the nonlocal case.
We describe in Sec. \ref{sec:two-qubit1} the main result of this Paper, namely how to measure a spin product observable on two qubits using a non-maximally entangled meter state.
We then compare it to the quantum erasure scheme in Sec. \ref{sec:compare}.
In Sec. \ref{sec:two-qubit2}, we study a possible alternative to the above using a maximally-entangled state and its impact on the post-measurement state of the system.
In Sec. \ref{sec:discussion}, we draw from the previous sections to establish a relation between measurement strength and ancillary entanglement in the two-qubit case. 
Finally, we conclude in Sec. \ref{sec:conclusions} by exposing the advantages and applications of this approach, as well as possible extensions.

\section{Generalized measurement of a single qubit}
\label{sec:one-qubit}

A VN generalized quantum measurement consists of an interaction between two quantum states, respectively called the system $S$, initially in the state $\ket{\psi}_S$, and a property of which we wish to measure ; and the meter $M$, prepared in a known initial state, which we will use to measure $S$. 
The interaction is followed by a projective measurement on $M$, in order to read out the result.
By designing an appropriate tunable interaction between the system and the meter, one can actually carry out measurements of different strengths, with much more flexibility than what is allowed by projective measurements.

\begin{figure}[htp]
    \includegraphics[width=0.40\textwidth]{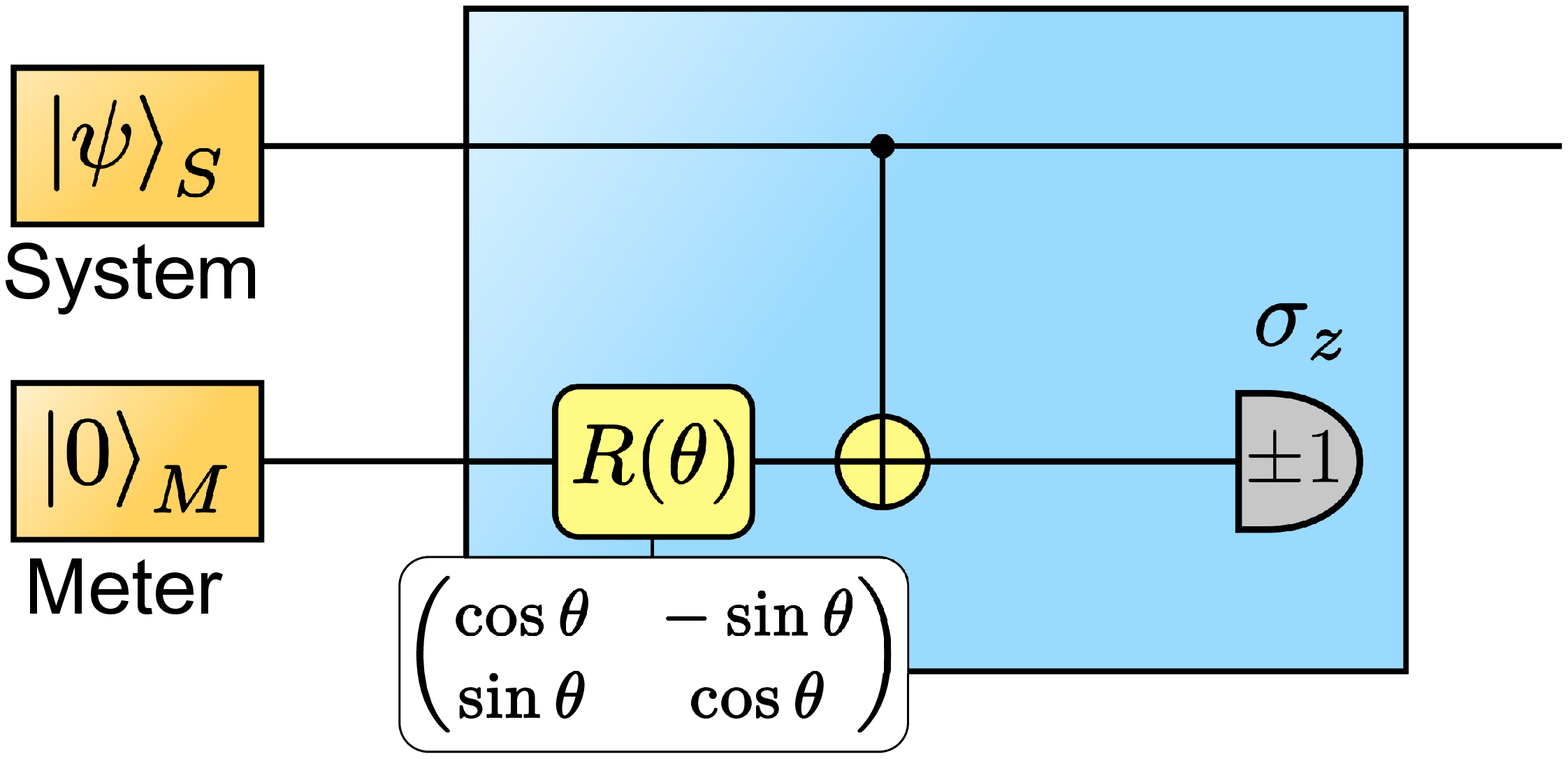}
    \caption{\label{fig:one-qubit}Quantum circuit representation of the one-qubit $\sigma_z$ spin indirect measurement model described in Sec. \ref{sec:one-qubit}.}
\end{figure}

Several such useful interactions have been proposed in the past for the measurement of single qubits (see \cite{Baek2008} for instance). 
We here focus on the one described in \cite{Lund2010}, that can be used to measure the system spin $\sigma_z$, and which is represented in \figref{fig:one-qubit}.
It consists in a local rotation applied to $M$ in order to obtain the following meter state:
\begin{align}
R(\theta)\ket{0}_M=\cos{\theta}\ket{0}_M+\sin{\theta}\ket{1}_M \label{eq:meter_1qubit}
\end{align}
followed by a Controlled-NOT (CNOT) gate between the meter and the system.

\addtocounter{figure}{1}

After, the result is retrieved via a projective measurement of $\sigma_z^M$ on $M$, the corresponding Positive-Operator Valued Measure (POVM) effects for the whole process are given by
\begin{align}
E_{\pm1}=\frac{1}{2}( \mathbb{1} \pm  \underbrace{\cos(2\theta)}_\text{strength}  \sigma_{z} ) \label{eq:povm_1qubit}
\end{align}
where $\mathbb{1}$ designates the identity operator.

Computing the statistics associated to this POVM reveals that the $\mathfrak{S}\equiv\cos(2\theta)$ factor acts as the \emph{measurement strength}, with $\mathfrak{S}=1$ corresponding to a strong measurement (perfect meter-system correlation) and $\mathfrak{S}=0$ corresponding to no measurement at all (no correlations):
\begin{align*}
    \mathfrak{S}&\rightarrow0 & P_{+1}&\rightarrow\frac{1}{2} & P_{-1}&\rightarrow\frac{1}{2}\\
    \mathfrak{S}&\rightarrow1 & P_{+1}&\rightarrow\left|\left<0|\psi\right>_S\right|^2 & P_{-1}&\rightarrow\left|\left<1|\psi\right>_S\right|^2    
\end{align*}
This generalized measurement scheme for one qubit has the advantage of being implementable using linear optics for polarization qubits \cite{Pryde2005} and has been used to test experimentally Ozawa's error-disturbance relations \cite{Ozawa2003,Baek2013,Edamatsu2016b} as well as to measure weak values \cite{Kaneda2014}.

\section{Generalized spin product measurement via a non-maximally entangled meter}
\label{sec:two-qubit1}

We consider a bipartite qubit system where a pair of qubits is distributed between Alice (A) and Bob (B).
For clarity, this pair of qubits is initially assumed to be in a pure (possibly entangled) state $\ket{\Psi}_S$.
Our goal is to answer the following: is it possible to extend the generalized measurement process of Sec. \ref{sec:one-qubit} described by the POVM in Eq. \eqref{eq:povm_1qubit} to the case of a two-qubit observable?

Namely, we will now attempt to extend our measurement of $\sigma_z$ to a measurement of the product observable $\sigma_{z_A}\sigma_{z_B}$.
It has already been shown that one can carry a projective measurement of $\sigma_{z_A}\sigma_{z_B}$ by using a maximally-entangled meter, e.g. the \emph{Bell state} $\ket{\Phi^+}_M$ \cite{Edamatsu2016a,Xu2019}. 
Following such previous approaches that established maximally-entangled qubit pairs (\textit{ebits}) as the standard resource for nonlocal quantum protocols, one may try to start with a nonlocal meter initialized in the state $\ket{\Phi^+}_M$.

A straightforward generalization of the process described in Sec. \ref{sec:one-qubit} would for instance consist in transforming this initial nonlocal meter state $\ket{\Phi^+}_M$ into a superposition of eigenstates associated with different outcomes, analogous to the one in Eq. \eqref{eq:meter_1qubit}: 
\begin{subequations}
  \begin{align}
  &\ket{\Phi^+}_M\rightarrow\ket{\Phi'}_M\equiv\cos{\theta}\underbrace{\ket{\Phi^+}_M}_\text{result $+1$}+\sin{\theta}\underbrace{\ket{\Psi^+}_M}_\text{result $-1$}\label{eq:meter_ideal}\\
  &=\frac{1}{\sqrt{2}}\left(\cos{\theta}\ket{00}+\sin{\theta}\ket{01}+\sin{\theta}\ket{10}+\cos{\theta}\ket{11}\right)\label{eq:meter_expanded}\\
  &=\cos\alpha\ket{++}+\sin\alpha\ket{--}\label{eq:meter_schmidt} 
  \end{align}
\end{subequations}
where $\ket{\Phi^+}_M$ and $\ket{\Psi^+}_M$ are the usual maximally-entangled Bell states, corresponding to global measurement outcomes $+1$ and $-1$ respectively, $\alpha=\frac{\pi}{4}-\theta$ and $\ket{\pm}\equiv\frac{1}{\sqrt{2}}\left(\ket{0}\pm\ket{1}\right)$.

However, interpreting Eq. \eqref{eq:meter_schmidt} as the Schmidt decomposition \cite{Nielsen2010} for the state $\ket{\Phi'}_M$ suggests that the transformation \eqref{eq:meter_ideal} is not realizable using only local unitary operations.
Eq. \eqref{eq:meter_schmidt} shows indeed that the meter state $\ket{\Phi'}_M$ is in general not maximally-entangled, hence not accessible from a Bell state via local unitaries \cite{Hulpke2006}. 

The state $\ket{\Phi'}_M$ can however be easily obtained from the state $\ket{\Phi^+}_M$ via some non-unitary operation that would discard unwanted amplitudes, in a fashion similar to a filter, in order to achieve the desired imbalance between the Schmidt coefficients of Eq. \eqref{eq:meter_schmidt}.  

Restricting ourselves to unitary operations, one can implement the transformation \eqref{eq:meter_ideal} probabilistically with a 50\% success rate, or deterministically using a classical communication channel between Alice and Bob as guaranteed by Nielsen's majorization theorem \cite{Nielsen1999}.
An example of such a possible implementation will be presented in Sec. \ref{sec:compare}.

In general, if one has an entangled qubit pair with known Schmidt coefficients $\lambda_0$ and $\lambda_1$, one can obtain such a state starting from the Schmidt basis and applying a Hadamard gate $H$ on each side.

\subsection*{Description of the measurement scheme}

Let us now assume that the non-maximally-entangled meter state $\ket{\Phi'}_M$ has been successfully prepared for some $\theta$ between 0 and $\frac{\pi}{4}$. 
Alice and Bob can now proceed to couple their qubits with the meter via local CNOT gates, as depicted in \figref{sfig:nmem}, before each (projectively) measuring their meter qubit.
For each of the four possible local outcomes, the final system state is given by the following measurement operators:
\begin{subequations}
    \label{eq:measop_1}
    \begin{align}
    \begin{split}\label{eq:measop_11}    
    M_{++}={}&M_{--} \\
            =&\frac{1}{\sqrt{2}} \left\{ \cos\theta \left( \Pi_{00} + \Pi_{11} \right) + \sin\theta \left( \Pi_{01} + \Pi_{10}  \right)\right\}  
    \end{split}\\
    \begin{split}\label{eq:measop_12}
    M_{+-}={}&M_{-+} \\
            =&\frac{1}{\sqrt{2}} \left\{ \sin\theta \left( \Pi_{00} + \Pi_{11} \right) + \cos\theta \left( \Pi_{01} + \Pi_{10} \right)\right\} 
    \end{split}
    \end{align}
\end{subequations}
where $\Pi_{ij}$ is the projector on $\ket{ij}$, i.e. $\Pi_{ij}=\ket{ij}\bra{ij}$.

\begin{figure}[htp]
    \includegraphics[width=0.45\textwidth]{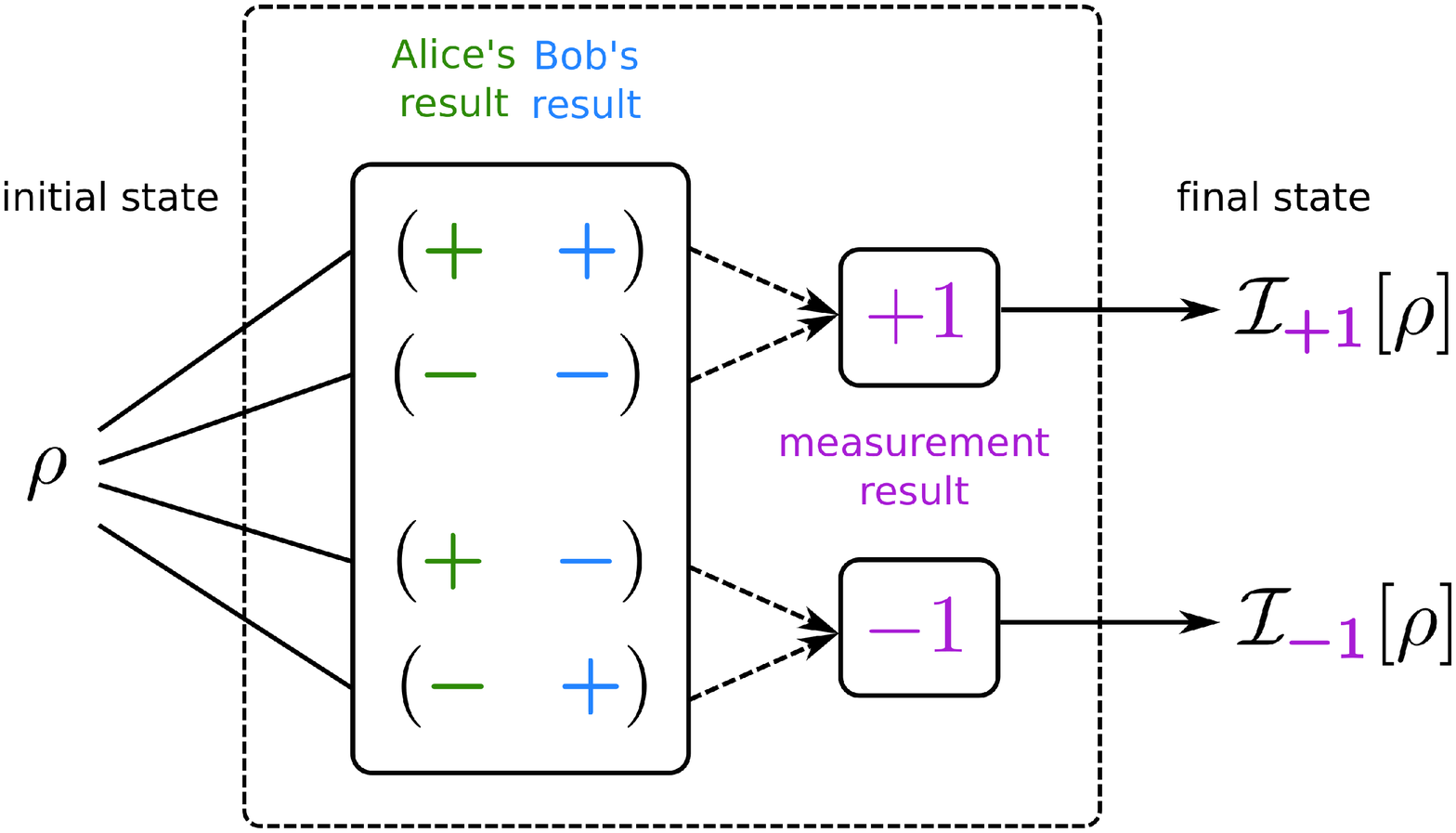}
    \caption{\label{fig:instrument}Schematic representation of the measurement process. 
    Once Alice's and Bob's results are multiplied together and any remaining local information is discarded, the measurement process is described by the quantum instrument $\mathcal{I}$.}
\end{figure}

From the four different local outcomes, the global outcomes are computed classically by allowing Alice and Bob to share their results.
Considering only the global outcomes and discarding any remaining local information, the evolution can be described by two different quantum operations, one for each result (see \figref{fig:instrument}).
The unnormalized post-measurement states of the system are given by the action of the following superoperators on the initial density matrix $\rho=\ket{\psi}\bra{\psi}$:
\begin{subequations}
    \label{eq:inst}
    \begin{align}
    \mathcal{I}_{+1}[\rho]&= M_{++} \rho M^\dagger_{++} + M_{--} \rho M^\dagger_{--} \label{eq:inst1}  \\
    \mathcal{I}_{-1}[\rho]&= M_{+-} \rho M^\dagger_{+-} + M_{-+} \rho M^\dagger_{-+} \label{eq:inst2}
    \end{align}
\end{subequations}
    
These operations form the \emph{quantum instrument} $\mathcal{I}$ \cite{Davies1970,Ozawa2004}, which fully encapsulates the measurement process as it provides a complete description of both post-measurement states and measurement statistics, as we will see below.

The POVM effects can be obtained directly from the quantum instrument $\mathcal{I}$, via the relation $E_r=\mathcal{I}_r^*[\mathbb{1}]$,
where * designates the superoperator adjoint, obtained by taking the adjoints of the measurement operators $M_{ij}$.
This yields: 
\begin{subequations}
    \label{eq:povm}
    \begin{align}
    E_{+1}&= M^\dagger_{++}M_{++} + M^\dagger_{--}M_{--} \label{eq:povm1} \\
    E_{-1}&= M^\dagger_{+-}M_{+-} + M^\dagger_{-+}M_{-+} \label{eq:povm2}
    \end{align}
\end{subequations}

Substituting with the expressions for the measurement operators \eqref{eq:measop_1}, the POVM can be rewritten in the following more compact way:
\begin{align}
E_{\pm1}=\frac{1}{2}\left( \mathbb{1} \pm \cos(2\theta) \sigma_{z_A}\sigma_{z_B} \right) \label{eq:povm3}
\end{align}

This is the desired nonlocal generalization of the POVM of Eq.\eqref{eq:povm_1qubit}, which yields the statistics expected from a genuine nonlocal measurement.

Moreover, we have $M_{++}=M_{--}$ and $M_{+-}=M_{-+}$, hence for a given global result, the evolution of the system does not depend on the local results. 
This allows us to rewrite the state evolution \eqref{eq:inst} in terms of two  \emph{effective measurement operators}, one for each global result:
\begin{subequations}
    \label{eq:effect_measop}
    \begin{align}
        M_+=& \cos\theta \left( \Pi_{00} + \Pi_{11} \right) + \sin\theta \left( \Pi_{01} + \Pi_{10}  \right) \label{eq:effect_measop1}\\
        M_-=& \sin\theta \left( \Pi_{00} + \Pi_{11} \right) + \cos\theta \left( \Pi_{01} + \Pi_{10} \right) \label{eq:effect_measop2}
    \end{align}
\end{subequations}

These operators only involve projectors on the two-dimensional eignespaces of the observable being measured, as is to be expected in the case of a degenerate observable, first studied by Luders \cite{Luders1951}.
All eigenstates thus remain unchanged by the measurement and this process is not entanglement-breaking, which are characteristics of an ideal nonlocal measurement.

This is the core result of this Paper: it is possible implement a nonlocal measurement of a spin product using only a meter state that need not be maximally-entangled. This is in sharp contrast with other nonlocal von Neumann measurement schemes developed so far \cite{Aharonov1986,Brodutch2016}.

\section{Comparison with the quantum erasure method}
\label{sec:compare}

\begin{figure}[htp]
  \includegraphics[width=0.47\textwidth]{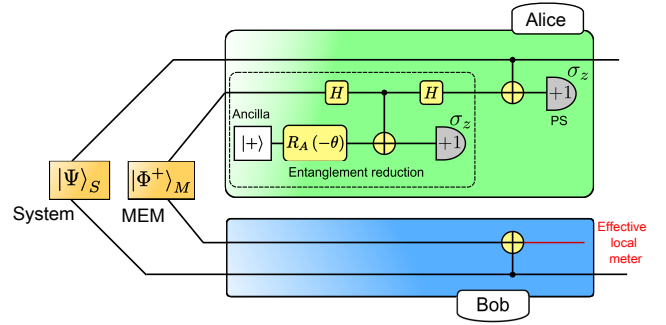}
  \caption{\label{fig:prepa}The entanglement reduction method: starting from a maximally-entangled meter (MEM), one first needs to reduce the entanglement using an additional local qubit before proceeding with the measurement process. Comparing this approach with the quantum erasure method of \figref{sfig:erasure} shows how the two measurement schemes are complementary in this particular case.  }
\end{figure}

The method we have just presented is deterministic, once the two parties are allowed to communicate. 
However if no communication between Alice and Bob is permitted whatsoever, Alice can still teleport her local result to Bob by post-selecting her part of the meter onto a known state. 
For causality reasons, this can only succeed with probability 50\%.
The result is then encoded in a single local meter on Bob's side.

We explained previously how to reduce the meter entanglement using non-unitary operations.
In this Section, we will however limit ourselves to unitary operations on each qubits, for comparison purposes with the protocol developed by Brodutch et al. \cite{Brodutch2016}, namely the quantum erasure method.
To this end, we consider the case where the two parties share a previously prepared maximally-entangled meter and are not allowed to communicate.

The quantum erasure method consists of four steps (see \figref{sfig:erasure}): first, a strong coupling between Alice's and Bob's systems and their shared maximally-entangled meter (MEM); followed by a post-selection on Alice's part of the MEM to teleport her result to Bob.
Then, Bob realizes a weak coupling between his remaining part of the MEM and an additional local meter.
Finally, Bob needs to erase the excess information contained in the MEM by projecting his part on the unbiased state $\ket{+}_{M_B}$.

In our scheme, Alice (or Bob) first implement tranformation \eqref{eq:meter_ideal} to reduce the entanglement of the meter, using for instance an additional ancillary local state (see \figref{fig:prepa}).
They subsequently proceed to strongly couple their systems with the resulting meter state.
The result can finally be teleported from one side to the other by post-selecting one part of the meter on a known state, say $\ket{0}_{M_A}$.

We thus show an example of a weak measurement without weak coupling \cite{Roik2019}: the weak coupling is replaced by a suitably prepared meter, in our case a non-maximally entangled meter.
The reduced entanglement guarantees that no excess information is stored in the meter, which makes the erasure step unnecessary.

\section{Generalized spin product measurement via a maximally entangled meter}
\label{sec:two-qubit2}

\begin{figure}[htp]
    \includegraphics[width=0.45\textwidth]{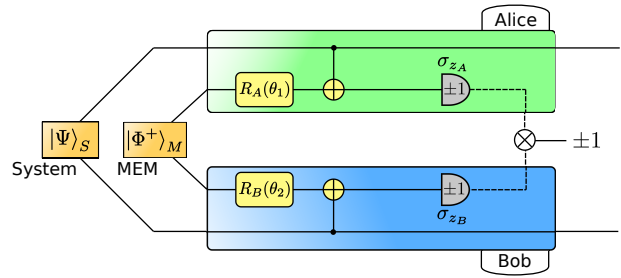}
    \caption{\label{fig:mem} Quantum circuit representation of the measurement described in Sec. \ref{sec:two-qubit2}.
    This time, Alice and Bob each applies a rotation and a CNOT interaction between their qubits and their shared maximally-entangled meter (MEM). }
\end{figure}

Before further discussing our results, it is interesting to study what might happen if we try to realize a nonlocal generalized measurement directly using a maximally-entangled meter, for instance the state $\ket{\Phi^+}$.
Instead of trying to achieve the transformation \eqref{eq:meter_ideal}, let us consider the meter state resulting from two local rotations implemented on Alice's and Bob's sides, of angles $\theta_1$ and $\theta_2$ respectively, as shown on \figref{fig:mem}. 

We obtain (up to a global phase) the following state:
\begin{align}
    \ket{\Phi^+}_M\xrightarrow[\, R_B(\theta_2)\, ]{\, R_A(\theta_1)\, }\frac{1}{\sqrt{2}}\big(\begin{aligned}[t]
                                          &\cos{\theta}\ket{00}-\sin{\theta}\ket{01}\\
                                          &+\sin{\theta}\ket{10}+\cos{\theta}\ket{11}\big)
                                                    \end{aligned}
                                                    \label{eq:meter_afterrotation} 
\end{align}
with $\theta\overset{\mathrm{def}}{=}\theta_2-\theta_1$. 
As expected, this is different from the state \eqref{eq:meter_expanded}; this will have consequences on the post-measurement system state. 

If Alice and Bob locally couple their meter qubits to their system qubits via CNOT gates and locally measure their meters (see \figref{fig:mem}), the corresponding measurement operators are:
\begin{subequations}
\label{eq:measop_2}
\begin{align}
M_{++} &= \frac{1}{\sqrt{2}} \left( \cos(\theta) \left( \Pi_{00} + \Pi_{11} \right) + \sin(\theta) \left( \Pi_{01} - \Pi_{10}  \right)\right) \label{eq:measop1}\\ 
M_{+-} &= \frac{1}{\sqrt{2}} \left( \cos(\theta) \left( \Pi_{01} + \Pi_{10} \right) + \sin(\theta) \left( \Pi_{00} - \Pi_{11} \right)\right) \label{eq:measop2} \\ 
M_{-+} &= \frac{1}{\sqrt{2}} \left( \cos(\theta) \left( \Pi_{01} + \Pi_{10} \right) - \sin(\theta) \left( \Pi_{00} - \Pi_{11} \right)\right) \label{eq:measop3} \\
M_{--} &= \frac{1}{\sqrt{2}} \left( \cos(\theta) \left( \Pi_{00} + \Pi_{11} \right) - \sin(\theta) \left( \Pi_{01} - \Pi_{10} \right)\right) \label{eq:measop4}
\end{align}
\end{subequations}

Using Eq. \eqref{eq:povm}, we obtain the same POVM as in Sec. \ref{sec:two-qubit1}:
\begin{align}
E_{\pm1}=\frac{1}{2}\left( \mathbb{1} \pm \cos(2\theta) \sigma_{z_A}\sigma_{z_B} \right) \label{eq:povm4}
\end{align}

However in this case, since $M_{++}\neq M_{--}$ and $M_{+-}\neq M_{-+}$, we see that a same global result can lead to two different state evolutions.
Indeed, some knowledge about the local state of the system can be retreieved from the phase information in the final state. 
Ignoring the individual outcomes (coarse-graining) thus adds classical noise to the system: the post-measurement state is in general mixed even if the initial state of the system was pure. 
Such a measurement process is sometimes labeled as an \textit{inefficient quantum measurement} \cite{Wiseman2009}. 

The amount of classical noise introduced by the coarse-graining can be evaluated via the difference in purity between the initial and the final states $\Delta\gamma$. 
It is found to be maximal when the initial state is an equal (in modulus) superposition of states associated with different global results, for instance $\ket{+}_A\ket{+}_B$.

In this case, the purity degradation $\Delta\gamma$ (going from an initially pure state $\gamma=1$ to a mixed state $\gamma<1$) can be related to the measurement strength $\mathfrak{S}$:
\begin{align}
    \Delta\gamma=\frac{1-\mathfrak{S}^2}{2}
\end{align}

We see that for a strong measurement ($\mathfrak{S}=1$) the system purity is unaffected, whereas for a weak measurement ($\mathfrak{S}\rightarrow0$), the system purity tends to $\frac{1}{2}$.

\section{Generalization and discussion}
\label{sec:discussion}

We saw previously that for a nonlocal generalized measurement to be efficient, i.e. without added classical noise, the entanglement of the meter state need to be adjusted in accordance with the desired measurement strength.
Hereafter, we shall use the concurrence \cite{Wootters1998} as our main measure of entanglement, defined as follows for a pure two-qubit state:
\begin{align}
    C\equiv2\lambda_0\lambda_1
\end{align}
where $\lambda_0$ and $\lambda_1$ are the Schmidt coefficients.

As was shown in Sec. \ref{sec:two-qubit1}, for a nonlocal measurement to be efficient, the meter state should be such that coefficients associated to same global outputs should be equal, as in Eq. \eqref{eq:meter_ideal}:
\begin{align}
    \frac{1}{\sqrt{2}}\left(\cos{\theta}\ket{00}+\sin{\theta}\ket{01}+\sin{\theta}\ket{10}+\cos{\theta}\ket{11}\right)    
\end{align}

It turns out that in this case, the resulting measurement strength $\mathfrak{S}$ is directly equal to the concurrence $C$ of the meter state:
\begin{align}
    C=\mathfrak{S}
    \label{eq:concurr_strength}
\end{align}

Let us now turn to the case when, as in Sec. \ref{sec:two-qubit2}, the entanglement $C$ contained in the meter state is higher than the desired measurement strength $\mathfrak{S}$.
It is then impossible to generate an ideal meter state, but one can still obtain the desired strength by applying appropriate local unitaries in order to prepare the following state:
\begin{align}
    \frac{1}{\sqrt{2}}\left(\cos{\theta}\ket{00}+e^{i\phi}\sin{\theta}\ket{01}+\sin{\theta}\ket{10}+\cos{\theta}\ket{11}\right)    
\end{align}
This is a generalized form of Eq. \eqref{eq:meter_afterrotation}.

The resulting phase $\phi$ is linked to the meter entanglement $C$ and the measurement strength $\mathfrak{S}$ by the relation:
\begin{align}
    \cos^2{\left(\frac{\phi}{2}\right)}=\frac{1-C^2}{1-\mathfrak{S}^2}    
    \label{eq:phi_relation}
\end{align} 
The ideal case of Sec. \ref{sec:two-qubit1} and the case of Sec. \ref{sec:two-qubit2} are recovered by setting $\phi=0$ and $\phi=\pi$, respectively.

The excess entanglement manifests itself through the added phase $\phi$, which in turn is responsible for the purity degradation of the post-measurement system state.
As in \ref{sec:two-qubit2}, this additionnal classical noise is maximal when the system being measured is initially in the state $\ket{+}_A\ket{+}_B$.  
The purity degradation can then be written as:
\begin{align}
    \Delta\gamma=\frac{1}{2}\left\{1-\left(\cos^2{\frac{\phi}{2}}+\mathfrak{S}\sin^2{\frac{\phi}{2}}\right)^2\right\}
    \label{eq:gamma_relation}
\end{align}
We recover the efficient measurement case ($\Delta\gamma=0$) by setting $\phi=0$ and the extreme noisy case of Sec. \ref{sec:two-qubit2} ($\Delta\gamma=\frac{1}{2}$) by setting $\phi=\pi$.

\begin{figure}[htp]
    \includegraphics[width=0.4\textwidth]{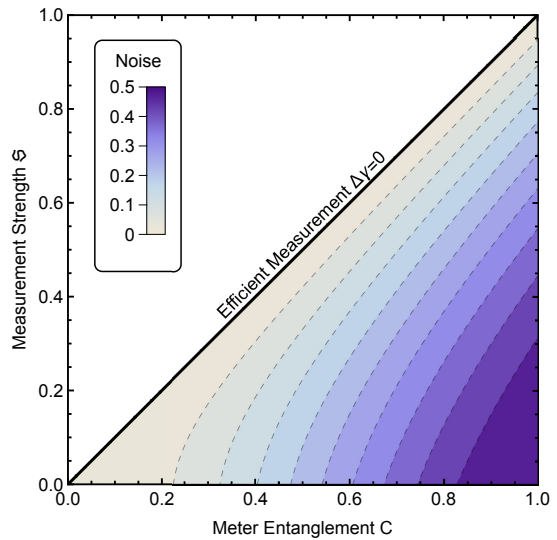}
    \caption{\label{fig:noise}Upper bound on the purity degradation $\Delta\gamma$ as a function of the meter concurrence $C$ and the measurement strength $\mathfrak{S}$. 
    The case $C=\mathfrak{S}$ of Eq. \eqref{eq:concurr_strength} is represented as a straight line, and corresponds to an efficient measurement with zero classical noise.} 
\end{figure}

One can combine relations \eqref{eq:phi_relation} and \eqref{eq:gamma_relation} to numerically evaluate the noise, as represented in \figref{fig:noise}.

We see that in order to make a measurement of strength $\mathfrak{S}$, one needs at least an amount of entanglement equal to $\mathfrak{S}$. 
A consequence of this fact is that a nonlocal strong measurement can only be achieved using Bell states.
We also notice that the noise increases non-linearly as the measurement strength deviates from the meter entanglement. 

\section{Conclusions}
\label{sec:conclusions}
In this Paper, we discussed a new approach to measure nonlocal spin products, using the formalism of generalized quantum measurements.
We found that one can achieve an efficient genuine nonlocal generalized measurement using a non-maximally entangled meter state.
In particular, we established relations between the desired measurement strength and the necessary entanglement for the measurement to be efficient, that is to say without any additional classical noise.
The effect of excessive entanglement was evaluated and found to be detrimental to the purity of the post-measurement state, but not to the overall measurement statistics.  
Another advantage of this new measurement scheme is that it does not require any quantum erasure step after the interaction. 
This approach is thus remarkably resource-efficient compared to other already existing schemes \cite{Brodutch2016,Kedem2010} and does not involve probabilistic steps.
It is also feasible using linear optics, using hyperentangled photon pairs for instance \cite{Xu2019}.

For clarity purposes, we focused our attention on the measurement of the spin product $\sigma_{z_A}\otimes\sigma_{z_B}$, but the proposed scheme can be easily adapted to measure any nonlocal spin product by applying appropriate one-qubit gates. 
Spin product measurement is a special case of nonlocal measurement as it is one of the few that can be directly measured in the von Neumann paradigm without violating causality.
Measuring spin products is crucial in tests of quantum nonlocality, such as testing Bell inequalities.
Measuring a spin product as been shown to be equivalent to measuring a modular sum, a relatively easier task. 
The question of whether or not our approach can be extended to more general observables remains open.

A promising application for this scheme resides in the measurement of weak values \cite{Aharonov1988}\cite{Laloe2012} in a nonlocal setting, which can be obtained directly as the weak limit of postselected conditioned averages \cite{Dressel2010}. 
Measuring nonlocal observables is also important in quantum error correction \cite{Gottesman1997} and variable measurement strength could be useful quantum computing without strong measurements \cite{Lund2011}.
 
\begin{acknowledgments}
    The authors wish to thank Aharon Brodutch for valuable discussions and Lev Vaidman for pointing us to relevant references. 
    This research was supported in part by JSPS KAKENHI Grant Number JP18J10639 and by MEXT Quantum Leap Flagship Program (MEXT Q-LEAP) Grant Number JPMXS0118067581. 
    P.V. thanks Tohoku University Division for Interdisciplinary Advanced Research and Education for their financial support. 
\end{acknowledgments}

%\bibliography{Bibliography.bib}

%merlin.mbs apsrev4-1.bst 2010-07-25 4.21a (PWD, AO, DPC) hacked
%Control: key (0)
%Control: author (0) dotless jnrlst
%Control: editor formatted (1) identically to author
%Control: production of article title (0) allowed
%Control: page (1) range
%Control: year (0) verbatim
%Control: production of eprint (0) enabled
%
         
\end{document}